\documentclass[11pt,twoside]{article}

\usepackage{asp2006}
\usepackage{epsf}
\usepackage{psfig}
\usepackage{lscape}

\begin{document}
\title{The Extremely Young Star Cluster Population In Haro 11}   
\author{Angela Adamo$^1$, G\"oran \"Ostlin$^1$, Erik Zackrisson$^1$, \& Matthew Hayes$^2$}  
\affil{$^1$ Department of Astronomy, Stockholm University, AlbaNova, SE-106 91 Stockholm, Sweden; $^2$ Observatoire Astronomique de l'UniversitŽ de Genve, 51, ch. des Maillettes,CH-1290 Sauverny, Suisse}    

\begin{abstract} 
We have performed a deep multi-band photometric analysis of the star cluster population of Haro 11. This starburst galaxy ($\log$ L$_{FUV} = 10.3$ L$_{\odot}$) is considered a nearby analogue of Lyman break galaxies (LBGs) at high redshift. The study of the numerous star clusters in the systems is an effective way to investigate the formation and evolution of the starburst phase. In fact, the SED fitting models have revealed a surprisingly young star cluster population, with ages between 0.5 and 40 Myr, and estimated masses between 10$^3$ and 10$^7$ solar masses. An independent age estimation has been done with the EW(H$\alpha$) analysis of each cluster. This last analysis has confirmed the young ages of the clusters. We noticed that the clusters with ages between 1 and 10 Myr show a flux excess in H (NIC3/F160W) and/or I (WFPC2/F814W) bands with respect to the evolutionary models. Once more Haro 11 represents a challenge to our understanding.
\end{abstract}


\section{Introduction}    
The interactions and/or mergers between galaxies are processes that usually enhance the star formation rate (SFR) (see other contributions in these proceedings) and produce many luminous star clusters (SCs). The cluster formation and evolution is strictly connected with the evolutionary history of the host galaxy. We have focused our studies on the local blue compact galaxy (BCG) Haro 11 (ESO 350-IG 038), with estimated distance of around 82 Mpc. It shows a complex morphology, with three active and very luminous starburst regions and irregular kinematics (\citealp{Ostlin01}, \"Ostlin et al. in prep.). A photometric study of the halo of Haro 11 \citep{micheva09} confirms that the stellar population in the outskirts of the system is consistent with a $\geq$10 Gyr old population and low metallicity. The coexistence of old stellar populations, active star forming regions \citep{Hayes07}, and unrelaxed kinematics indicates that the Haro 11 starburst could be the result of a merger between an evolved system and a gas-rich dwarf galaxy. As a consequence of this merging a new population of SCs has been formed. In Figure 1 we show the image of the galaxy in the WFPC2 filter, F606W. The starburst region is dominated by the numerous SCs. Our aim is to investigate the SC properties to constraint the starburst age and propagation.
\begin{figure}[ht]
\begin{center}
\scalebox{0.5}{\rotatebox{0}{\includegraphics{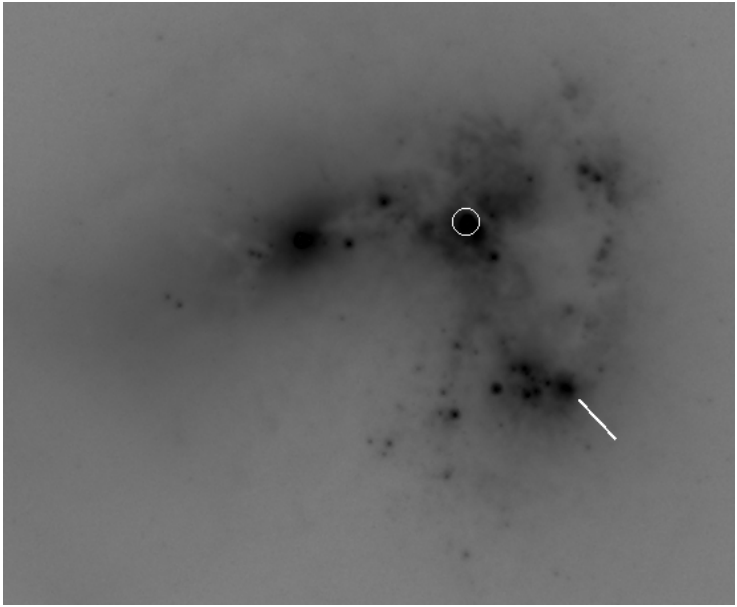}}}
\end{center}
\caption[]{WFPC2/F606W image of Haro11. The galaxy shows three starburst regions with numerous clusters. The central knot, where a white circle is located appears obscured by some dust lines. Inside the circle there is the brightest cluster of Haro 11. The line shows the position of the brightest young cluster.}
\end{figure}

\section{Multi-wavelength analysis} 

A multi-wavelength photometric analysis (from far-UV to IR), have been performed on the SCs. About 300 confirmed cluster candidates have been detected in the high-resolution HST images (see \citealp{adamo09}). We show here the results of the SED (spectral energy distribution) analysis (Adamo et al., in prep.). The used models are based on the \citet{Erik01} population synthesis code, which includes nebular emission from the photoionized interstellar medium (important when analyzing observations of young stellar populations; e.g. \citealp{Anders03}; \citealp{Erik08}). An instantaneous burst, a Salpeter IMF and a metallicity of Z$= 0.004$ has been used as input. In Figure 2 (left panel), we show an example of the fit. When we try to fit the H and I bands we obtain a poor fit of the blue side of the spectrum (more sensitive to age). In this case we obtained an age of 35 Myr. If we only fit the wavelength range from UV (ACS/F140LP) to V (WFPC2/F606W), excluding H and I, we get a better fit of the blue side. The age estimated in this second case is of 3.5 Myr.  Comparing the two different spectral models with the observed fluxes (Figure 2) we see a clear excess in H and I bands that quite considerably influences the age estimation (and masses) of the SCs. In order to get an independent estimation of the ages of the SC we used also EW(H$\alpha$) Ð a sensitive estimator at young ages. However, we have to consider the ages estimated from H$\alpha$ as an upper limit to the real ages of the clusters. In fact, we don't know how many photons escape from their birth place or are lost because of the way we estimate the flux on the cluster positions (we used a fixed aperture radius, optimized with respect to the PSF of the image). Moreover, in crowded regions we have to take into account contaminations from the youngest clusters against the more old and fainter H$\alpha$ emitters. Figure 2 (right panel) shows a comparison between the ages estimated from SED fit (H and I included) and EW(H$\alpha$). The ages estimated from EW(H$\alpha$) are between 1 - 35 Myr, much lower than the SED fit outputs (1 - 100 Myr). Because of the unclear origin of this red excess at the youngest ages \citet{Reines08} we canÕt produce any ad-hoc model to fit these young clusters. Finally we decided to exclude the H and I from the chi square fits.
\begin{figure}[!ht]
\plottwo{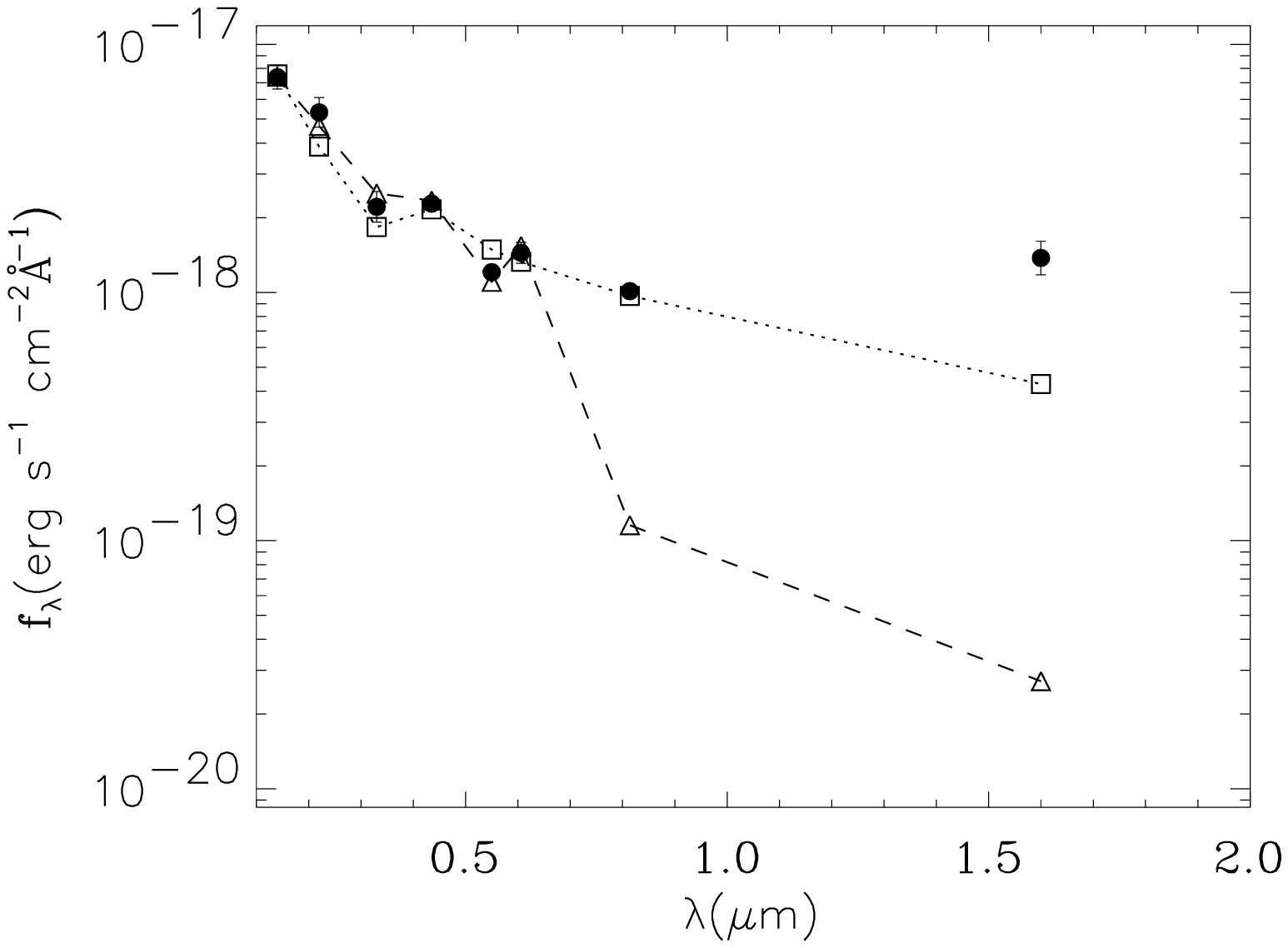}{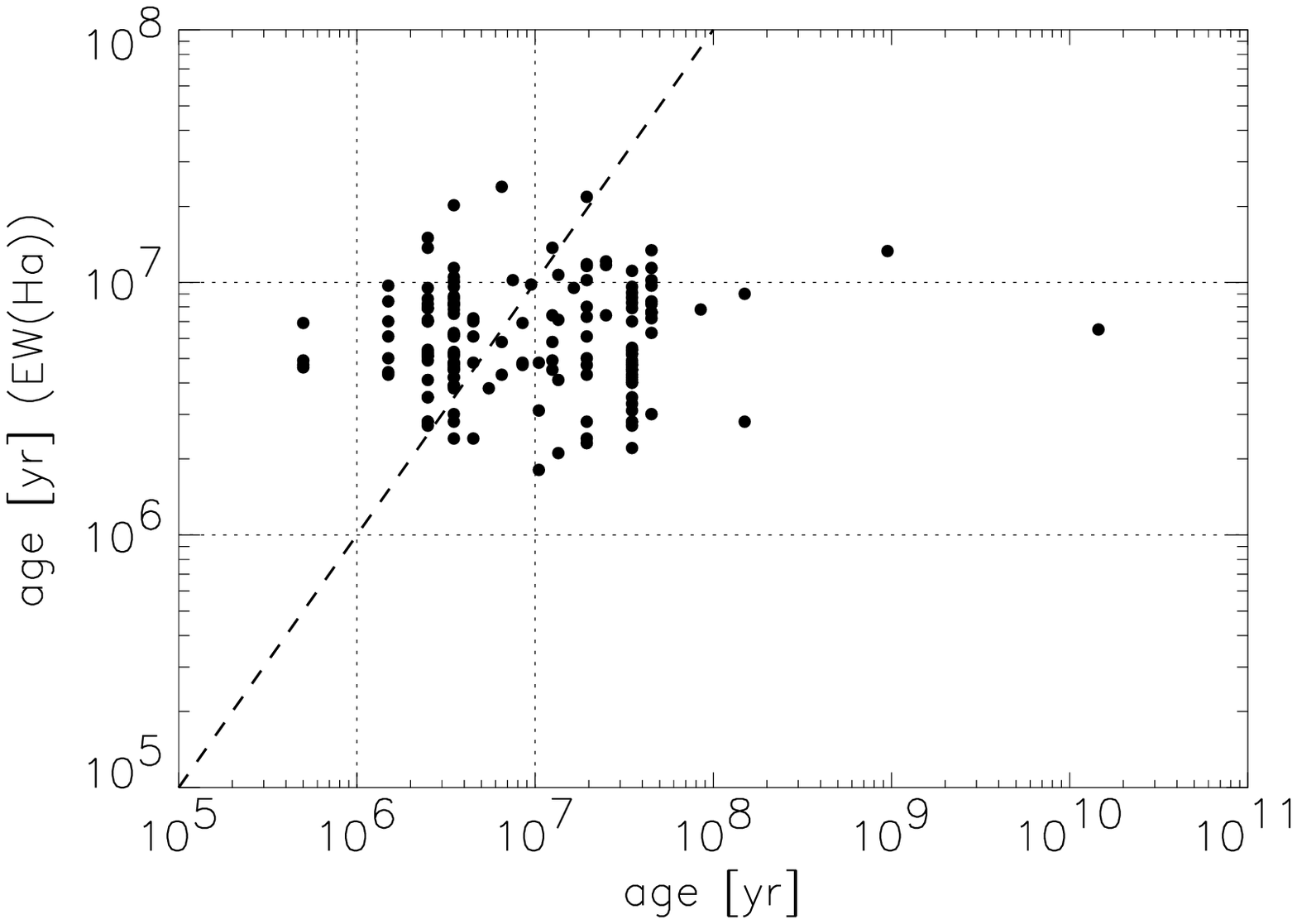}
\caption{Left plot: the best SED fit to the detected fluxes (filled dots). The squares connected by a dotted line show the best fit model to all the filters. The triangles connected by a dashed line is the best fit to the blue side of the spectrum, excluding the I and H bands from the fit. The red excess is seen as an excess in flux at the redder wavelength with respect to the best fit model. Right plot: the cluster age estimated from EW(H${\alpha}$) versus the ages obtained from the SED fit of all the detections (I and H included). The dashed line   indicates the region of the plain where the ages estimated with the two method agree. The dotted lines outline the region of the plain where the cluster ages is between 1 and 10 Myr. } 
\end{figure} 
In Figure 3 we can see the present mass as function of the ages of the SCs in Haro11. The clusters appears young, with ages between 1 and 40 Myr and masses between 10$^3$ and 10$^7$ M$_{\odot}$. 
\begin{figure}[ht]
\begin{center}
\scalebox{0.4}{\rotatebox{0}{\includegraphics{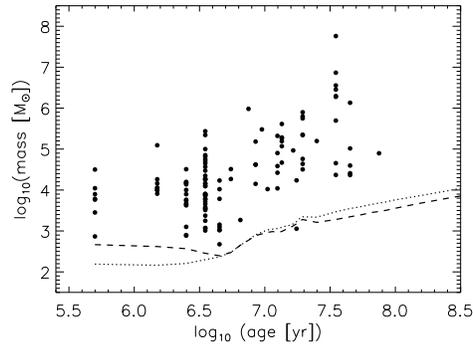}}}
\end{center}
\caption[]{The present mass as function of age of the clusters. The lines are the predicted detection limits using our stellar evolutionary models, for M$_V$=28.0 mag (dotted) and M$_I$=28.0 mag (dashed).}
\end{figure}
 
 \section{SFR from the cluster luminosity} 
We tried to estimate the evolution of the star formation rate (SFR) in Haro 11, using the relation between the SFR and the luminosity of the brightest young cluster in the galaxy \citet{Bastian08}. From the estimated Haro 11 far infrared  luminosity, L$_{FIR}$ \citet{Hayes07}, we know that the present SFR is around 18 M$_{\odot}$ yr$^{-1}$. Applying the relation by \citep{Bastian08}, we obtain from the brightest young cluster (M$_V=-14.5$ mag and age of 1.5 Myr, indicated by a line in Figure 1) a current SFR of 18.5 M$_{\odot}$ yr$^{-1}$, in perfect agreement with FIR estimations. However, the brightest cluster in the galaxy has M$_V=-15.9$ mag and an age of 35 Myr (inside the circle in Figure 1). For this cluster the previous relation gives a SFR, at the time when this cluster has formed, of 105 M$_{\odot}$ yr$^{-1}$. This suggests that after a first effective burst the star formation has propagated in the galaxy with less efficiency possibly caused by feedbacks and outflows. 

\section{Conclusion}
The analysis of the SCs in Haro 11 has confirmed the young age of the ongoing starburst in the system. The starburst involves clusters with mass range of several orders. The SED analysis shows that a subsample of clusters have a significant flux excess in H and I with respect to the predicted stellar population models. The origin of this red excess is not still clear and is impossible to match with the usual stellar evolutionary models used so far. We will discuss our results in forthcoming papers.

\acknowledgements 
A. Adamo thanks the organizers for the interesting conference. We acknowledge support from the Swedish Research Council (Vetensapsr{\aa}det).


\end{document}